# Selection committees for academic recruitment: does gender matter?[1]

Giovanni Abramo[a,*], Ciriaco Andrea D'Angelo[b,a], Francesco Rosati[c]

[a] *Laboratory for Studies of Research and Technology Transfer*
*Institute for System Analysis and Computer Science (IASI-CNR)*
*National Research Council of Italy*
ADDRESS: Consiglio Nazionale delle Ricerche
Istituto di Analisi dei Sistemi e Informatica
Via dei Taurini 19, 00185 Roma – ITALY
tel. and fax +39 06 72597362, giovanni.abramo@uniroma2.it

[b] *Department of Engineering and Management*
*University of Rome "Tor Vergata"*
ADDRESS: Dipartimento di Ingegneria dell'Impresa
Università degli Studi di Roma "Tor Vergata",
Via del Politecnico 1, 00133 Roma – ITALY
tel. and fax +39 06 72597362, dangelo@dii.uniroma2.it

[c] *Department of Management Engineering*
*Technical University of Denmark*
ADDRESS: Technical University of Denmark
Produktionstorvet Building 426
2800 Kgs. Lyngby - Denmark
tel +45 45256021, frro@dtu.dk

**Abstract**

Underrepresentation of women in the academic system is a problem common to many countries, often associated with gender discrimination. In the Italian academic context in particular, favoritism is recognized as a diffuse phenomenon affecting hiring and career advancement. One of the questions that naturally arises is whether women who do assume decisional roles, having witnessed other phenomena of discrimination, would practice less favoritism than men in similar positions. Our analysis refers to the particular case of favoritism in the work of university selection committees responsible for career advancement. We observe a moderate positive association between competitions with expected outcomes and the fact the committee president is a woman. Although committees presided by women give more weight to scientific merit than those presided by men, favoritism still occurs. In fact, in the case the committee president is a woman, the single most important factor for the success of a candidate is joint research with the president; while in the case of male presidents, it is the years together in the same university.

# 1. Introduction

In the current economic system, the competitiveness of organizations depends above all on their capacity to attract and select the best talent available on the labor market. Phenomena of *a priori* discrimination (gender, age, race, religion, etc.) represent obstacles to the operation of effective and efficient production systems. To discard a potentially talented candidate on the basis of their demographic, socio-cultural or other characteristics represents both an opportunity cost for the organization and a gain for competitors.

Avoidance of discrimination in the higher education sphere is still more important, given the role that universities play in support of industrial competitiveness, socio-economic development and social mobility. In competitive education systems diversity is actually seen as an advantage, and universities compete with one another to bring in the best researchers and teaching professors possible from both at home and abroad. Such competitive mechanisms have generally developed in a natural manner, supported by national policies stimulating the birth and development of true markets of higher education. In competitive systems, discriminatory phenomena are thus inhibited *a priori*. In contrast, the extensive state control that characterizes many European higher education systems has inhibited the activation of true competitive mechanisms. In these systems, recruitment and advancement take place through regulated procedures, fixed at a central level. Such systems provided for centralized competitions intended to guarantee the choice of the best among all incoming candidates and then provide for step by step advancements through their careers. However the theory behind such approaches is in reality seen to have various problems. For example in Italy, where recruitment and career advancement occur exclusively through national public competitions, common professional and public opinion is that merit is not the prevalent criteria for selection. It is in fact quite common to see systemic problems and dramatic individual cases discussed in the press, or even taken before the courts (Perotti, 2008; Zagaria, 2007). Problems of this sort are not limited to Italy alone: international scientific literature has dedicated significant space to the study of academic recruitment and promotion, including inquiries into gender and minority discrimination (Danell and Hjerm, 2013; Zinovyeva and Bagues, 2012; van den Brink et al., 2010; Stanley et al., 2007; Cora-Bramble, 2006; Price et al., 2005; Trotman et al., 2002). One line of this literature demonstrates that discriminatory phenomena tend to appear when evaluations are not based on transparent criteria (Rees, 2004; Ziegler, 2001; Husu, 2000; Ledwith and Manfredi, 2000; Allen, 1988). In effect, the academic recruitment is often described as an informal process, in which a few powerful professors promote or select new professors through mechanisms of cooptation (van den Brink et al., 2010; Husu, 2000; Fogelberg et al., 1999; Evans, 1995). Such mechanisms often disguise phenomena of personal favoritism, however this aspect of the problem has only been studied in a few countries, such as Turkey (Aydogan, 2012), Australia (Martin, 2009), Spain (Zinovyeva and Bagues, 2012) and Italy (Abramo et al., 2014a; Perotti, 2008; Zagaria, 2007). In examining the Spanish academic system, Zinovyeva and Bagues (2012) concentrate on the role of the connections between candidates for faculty promotion and the evaluators composing the examining boards. They show that the future performance of candidates who were promoted and had a weak connection with the evaluators was better than that of the other promoted candidates. Conversely, the performance of successful candidates with a strong link to the evaluators is worse than that of other promoted candidates, both before and after their promotion.

In a preceding work (Abramo et al., 2014b), we evaluated the efficiency of the selection process for career advancement of Italian university professors, specifically referring to all competitions for associate professor posts (1,232) in the national system for the year 2008. The analyses showed that in the three years following the competitions, the new associate professors were on average more productive than their incumbent colleagues. However several critical issues appeared. One of these concerned repeated cases of candidates who were rejected by the selection committees but then outperformed the winners in research productivity over the subsequent triennium. Another issue concerned cases of competition winners who resulted as totally unproductive. The analyses of the selection outcomes of each committee showed that almost half of them selected candidates who would go on to achieve below-median productivity in their field of research over the subsequent period.

In a subsequent study (Abramo et al., 2014c), we investigated the determinants of the results emerging from the first one (Abramo et al., 2014b). Our intention was to interpret the factors that could have contributed to the outcomes of the 2008 round of academic recruitments. In particular, we found that favoritism strongly conditioned the selection outcomes: the fundamental determinant of a candidate's success was not his or her scientific merit, rather the number of their years of service in the same university as the committee president. In addition, the fact that some committees were composed of members of higher scientific quality did not lead to less favoritism. Finally, there were no significant differences in favoritism practices between geographic areas.

In this work we now explore aspects of gender influences on academic selection processes. Scientific debate has brought attention to a series of factors that contribute to broad problems in the role of women in the academic system, present in many countries, from low percentages of women academics to outright discrimination affecting career opportunities. Although there are more female than male undergraduate and graduate students in many countries (OECD, 2014), female professors are under-represented in many faculties, particularly in senior positions (Moss-Racusin et al., 2012), They also progress more slowly through academic ranks, and have not attained important leadership roles (Rotbart et al., 2012; Bilimoria and Liang, 2011; Ceci and Williams, 2011; McGuire et al., 2004; Wright et al., 2003). Finally, despite improvements women professors earn less than men in comparable positions, and undergo funding disparities as well (Shen, 2013). The fact that women are underrepresented in decision-making positions and as reviewers (Wennerås and Wold, 1997) may lead to the conclusion that probabilities for the recruitment and advancement of female candidates are reduced. As a matter of fact, the available empirical evidence shows contradictory results. In examining the rating of proposals from the National Science Foundation, Broder (1993) finds that female reviewers rate female-authored proposals lower than do male reviewers. Bagues et al. (2014) estimated the causal effect of committees' gender composition in the 2012 Italian habilitation to full and associate professor positions. They found that each additional female evaluator decreases by 2 percentage points the success rate of female candidates. In examining 1539 career grant applications, Van Arensbergen et al. (2013) find no correlation between the number or percentage of women in a panel and the gender bias in the results. Conversely, Foschi and Valenzuela (2008) investigating hiring decision for junior engineer positions, find that in the choice of applicant there is no bias against the female candidate by male assessors and that there is a bias in her favor by female assessors. De Paola and Scoppa (2014) find that "female candidates to professor positions in Italy are less likely to be promoted when the committee is composed exclusively by males, while the gender gap disappears when the candidates are evaluated by a mixed sex committee". Similarly, Zinovyeva and Bagues (2012; 2011) find that in competitions for full professor positions in Spain, evaluators tend to favor candidates who belong to their own academic network and are also of the same gender. Van den Brink (2009) also suggests that gender bias in recruitment procedures in the Netherlands is due to the composition of panels.

In Italy, starting from a situation of low intake and given the presence of further discriminatory and negative phenomena, women are under-represented in senior faculty and decisional roles. One of the questions that naturally arises is whether the women who do assume decisional roles, having witnessed phenomena of discrimination, would practice less discriminatory behavior than men in similar positions. In a context where favoritism in public-sector and university employment competitions is diffuse (such as in Italy), our precise question is if women are more objective than men, and so less inclined to favoritism in candidate evaluations. The implications of such a question clearly extend to policy considerations, for example in determining an adequate representation of women in committees for candidate evaluation, and as committee president in particular.

The specific objectives of the current study are: i) to examine the association between the gender of the president and members of the selection committees and the expected outcome of academic recruitment; and ii) to analyse the different influences of the factors which may determine the selection outcome when the committee is respectively presided by a man or woman. The two potential determinants of interest are i) the scientific merit of the candidates, and ii) the possibility for favoritism (including nepotism), in terms of social proximity and research collaboration between the candidates and their evaluators, particularly the committee president.

The next section of the paper describes the structure of the Italian higher education system, particularly the measures adopted in 2008 for recruitment of associate professors. Sections 3 and 4 present the methodology and dataset used for the analyses. In section 5 we present the results of the tests of association conducted between applicant gender, gender of the evaluators and outcome of the competitions, followed by the results of regression analyses. The work concludes with the authors' discussion.

**2. The recruitment process in Italian universities**

The Italian Ministry of Education, Universities and Research (MIUR) recognizes a total of 96 universities as authorized to issue degrees. Sixty-seven of these are public universities, employing around 95% of all Italian faculty members.

In keeping with the so-called Humboldtian model of university policy, there are no "teaching-only" universities in Italy. All professors are required to carry out both research and teaching. Legislation includes a provision that each faculty member must provide a minimum of 350 hours of teaching per year. All new personnel enter the university system through public competitions, and career advancement can only proceed by further public competitions. Salaries are regulated at the centralized level and are calculated according to role (administrative, technical, or professorial), rank within role (for example assistant, associate or full professor) and seniority. None of a professor's salary depends on merit. Moreover, as in all Italian public administration, the dismissal of unproductive employees is unheard of.

The recruitment and advancement of professors is regulated by laws, overseen by the MIUR. There have been major reforms over recent years. Law 240 of 2010 introduced a double evaluation procedure for the selection of associate and full professors. The first level is a national habilitation of candidates, managed directly by the MIUR. The second set of evaluations is managed by the individual universities, to then choose the habilitated individuals best suited to the specific needs of each institution. Prior to Law 240, the processes of recruitment and career advancement were in the hands of the individual universities, following procedures dictated at the central level. The last major competition under the old system was held in 2008.

In the Italian university system all professors are classified in one and only one field (Scientific Disciplinary Sector or SDS, 370 in all), grouped into disciplines (University Disciplinary Areas or UDAs, 14 in all)[2]. In both the new and old system, competitions for recruitment and advancement are organized at the SDS level. The 2008 competition procedures required appointment of committees to judge the curricula of the candidates. Each committee was to be composed of five full professors belonging to the SDS for which the position was open. One member, the president, was designated by the university holding the competition and the other four were drawn at random from a short list of other full professors in the national SDS. The short list was in turn established by a vote of all full professors in the SDS.

The task of each committee was to provide a judgment of all candidates based on examination of the candidates' documented qualifications, and name at most two winners. The university announcing the competition could then hire one of the two top finishers. The other top finisher remained eligible for hiring by any other university in the national system without further competition, at any time over the next five years.

In order to rationalize the process of the individual competitions over the entire system, the MIUR monitored and gathered the hiring proposals of the various universities and supported the evaluation procedures through information management systems aimed at better guaranteeing transparency. One of the ministry measures was to provide a Web portal[3] with all the basic information on the competition procedures, the posts available, the number of candidates for each competition, the scheduling of the procedures and final results (winners list, etc.). The transparency provisions, the nomination of a national committee of experts in the field, and the timely issue of regulations for the evaluation procedures were all intended to ensure efficiency in the selection process. In reality, the characteristics of Italian system – such as the generally strong inclination to favoritism, the structured lack of consequences for poor performance by research units, and the lack of incentive schemes for merit – undermined the credibility of selection procedures for hiring and advancement of university personnel, just as happens for the Italian public administration in general. This is demonstrated by the growing number of legal cases brought by losing candidates and by published studies of systemic problems (Perotti, 2008; Zagaria, 2007). Also, previous bibliometric analysis has demonstrated that in university selection procedures, the most significant of several determinants of candidates' success is not scientific merit, rather the number of years that the selection committee president and the candidate have belonged to the same university, and whether the applicant and president have previously cooperated in research work (Abramo et al., 2014c).

---

[2] For the complete list see http://attiministeriali.miur.it/UserFiles/115.htm, last accessed on February 16, 2015.

[3] The MIUR Web portal, titled "Comparative evaluation in the recruitment of University Professors and Researchers (Law 3, 3 July 1998, no. 210)", is at http://reclutamento.murst.it/ last accessed on February 16, 2015.

## 3. Methods and data

### 3.1 Measuring scientific merit

In the current study we measure the scientific merit of a candidate in terms of his or her research productivity, embedding both quantity and quality of output. We acknowledge that research performance is not the only dimension of merit of a candidate. Evaluations of applicants should also consider dimensions representing the other two institutional missions of universities, meaning teaching and technology transfer. Furthermore, the assessment of research performance by quantity and quality of output alone neglects other attributes of the scientists' activities, for example the ability to manage research teams, to attract funds, their activities in consulting, editorial work, and so on. We would expect some level of correlation between research productivity and these other variables, however caution is recommended in the interpretation of our results, in which we consider only dimensions involving research output.

This subsection presents the methodology for the measurement of research productivity in those disciplines where the prevalent form for codifying research output is publication in scientific journals. As a proxy of total output, in this work we consider only the document types (articles, article reviews, and proceeding papers) indexed in the Web of Science (WoS), and adopt the fractional counting method. Because citation behavior varies by field, we standardize the citations for each publication to the average of the citation distribution for all the Italian cited publications of the same year and the same WoS subject category[4]. Finally, we account for the differences in the publication intensity across fields, comparing performance of individuals in the same field.

For professors of the same academic rank[5], we measure each individual's average yearly productivity by applying the indicator Fractional Scientific Strength (FSS)[6]:

$$FSS = \frac{1}{t}\sum_{i=1}^{N}\frac{c_i}{\bar{c}}f_i$$

[1]

Where:
t = number of years of work of the researcher in the period under observation;
N = number of publications of the researcher in the period under observation;
$c_i$ = citations received by publication *i*;
$\bar{c}$ = average of distribution of citations received for all cited publications indexed in the

---

[4] Abramo et al. (2012a) demonstrated that the best-performing scaling factor is the average of the distribution of citations received for all cited publications of the same year and subject category.
[5] In Italian universities higher academic ranks are more productive than lower ones (Abramo et al., 2011).
[6] A more extensive theoretical discussion of how to operationalize the measurement of productivity is found in Abramo and D'Angelo (2014).

same year and subject category of publication *i*;
$f_i$ = fractional contribution of researcher to publication *i*.

Fractional contribution $f_i$ is the inverse of the number of authors, in scientific fields where the practice is to place the authors in simple alphabetical order. However in the life sciences, widespread practice is to indicate the levels of individual contribution to the research by the order of names in the byline. Thus for these SDSs, we assign fractional weights according to the order of authors in the byline and whether the co-authorship involves intra-mural or extra-mural participation (see Abramo et al., 2013). If first and last authors belong to the same university, 40% of citations are attributed to each of them; the remaining 20% are divided among all other authors. If the first two and last two authors belong to different universities, 30% of citations are attributed to first and last authors; 15% of citations are attributed to second and last author but one; the remaining 10% are divided among all others[7].

We extract the data on the faculty of each university and their SDS classifications from a database on university personnel maintained by the MIUR. The bibliometric dataset is extracted from the Italian Observatory of Public Research (ORP), a database developed and maintained by the authors and derived under license from the Thomson Reuters WoS. Beginning from the raw data of the WoS, and applying a complex algorithm for reconciliation of the author's affiliation and disambiguation of the true identity of the authors, each publication (article, article review and conference proceeding) is attributed to the university scientist or scientists that produced it (D'Angelo et al., 2011). With the authorship of the publications thus clearly assigned we are able to calculate rankings of research productivity at the individual level and on a national scale. Based on the values of FSS for each scientist, for each SDS and academic rank, we obtain a national ranking list expressed on a percentile scale of 0-100 (worst to best).

## 3.2 Data

In 2008, 74 universities announced a total of 1,232 competitions for associate professor positions, concerning 299 SDSs. On average, they took two years to complete. At the moment of our data observation, 1,162 of the 1,232 competitions were concluded, resulting in the naming of 2,195 winners[8] out of the 16,500 candidates.

The ratio of competition winners to the numbers of the existing national associate-professor faculty averages 12.0%. In five UDAs the number of applications was higher than the current number of tenured associate professors. In Industrial and information engineering there were 2,010 applicants compared to the 1,493 active associate professors.

To ensure the representativeness of publications as proxy of research output, our bibliometric analysis focuses only on competitions launched in those SDSs where at least 50% of Italian professors produced at least one publication indexed in the WoS over the period 2004-2008. For brevity, we refer to these SDSs as to the *hard sciences*. Applying these criteria, we identify 654 competitions in a range of 193 SDSs for potential inclusion in the analysis. The only way to identify all the applicants in these

---

[7] The weighting values were assigned according to advice from senior Italian professors in the life sciences. The values could be changed to suit different practices in other national contexts.
[8] Each competition announced two winners, except for 39 that announced only one winner.

654 competitions would be to read the minutes of each selection committee, typically published on-line by the individual universities. Given the prohibitive scope of such a task we randomly extracted a further subset of 287 competitions (44% of the preliminary field of 654). Within this subset, the winners (550 in all) represent 22% of the total competition candidates (2,590). Candidates who were incumbent assistant professors experienced a more favorable rate of selection (23%) than candidates holding no faculty position (6.5%). Due to our methodology for solving ambiguity in author names, we can only measure the productivity of applicants who are listed as university faculty members. Thus our analysis of career advancement concentrates solely on the 2,314 candidates already in faculty role. In addition, for the measure of research productivity to be robust, it must be calculated over a sufficiently long period (Abramo et al., 2012b). Because of this, the analysis excludes assistant professors who entered faculty less than three years prior to the date of the competition.

The final dataset for the analysis is thus composed of 1,979 assistant professors, 473 of which were competition winners. Table 1 presents the characteristics of the candidate dataset by UDA and the coverage of the competitions relative to the total 193 hard science SDSs (124 SDSs are represented).

Table 2 shows the composition of the selection committees by gender for the 1,162 competitions that had concluded their deliberations as of the date of observation, in all SDSs. The table shows that only 16.1% of competitions (187 of 1,162) had a woman committee president. The last column of the table further shows that for the SDSs involved in these competitions, female full professors composed 18.7% of the total full professor faculty (3,361 of 17,985). These data confirm the under-representation of women in the entire Italian academic faculty composition. However the table also shows that the percentage of women called to preside over the selection committees is still lower. Also, 45.2% of the committees do not include any female members; 29.9% have one female professor; 16.4% have two; 6.3% have three, while 1.6% have four and 0.6% of committees are entirely women,.

The data confirm the results of many other studies on female underrepresentation in decisional and leadership roles in the academic sphere (e.g. Wright et al., 2003; Rotbart et al., 2012).

| UDA | Competitions | SDSs concerned | Winners | Academic winners with seniority ≥ 3 years |
|---|---|---|---|---|
| Mathematics and computer science | 26 (46%) | 7 (78%) | 50 (46%) | 45 (47%) |
| Physics | 19 (42%) | 5 (63%) | 37 (43%) | 30 (41%) |
| Chemistry | 25 (46%) | 8 (67%) | 47 (46%) | 44 (48%) |
| Earth sciences | 6 (30%) | 4 (33%) | 10 (27%) | 5 (17%) |
| Biology | 25 (34%) | 14 (74%) | 49 (34%) | 39 (31%) |
| Medicine | 62 (41%) | 32 (68%) | 116 (40%) | 87 (40%) |
| Agricultural and veterinary sciences | 15 (31%) | 11 (39%) | 27 (29%) | 26 (30%) |
| Civil engineering and architecture | 11 (42%) | 6 (86%) | 22 (43%) | 22 (46%) |
| Industrial and information engineering | 86 (60%) | 31 (74%) | 170 (62%) | 155 (62%) |
| Pedagogy and psychology | 5 (24%) | 3 (60%) | 8 (20%) | 7 (21%) |
| Economics and statistics | 7 (39%) | 3 (75%) | 14 (39%) | 13 (41%) |
| Total | 287 (44%) | 124 (64%) | 550 (43%) | 473 (44%) |

*Table 1: Population subset selected for analysis (in parentheses the percentage with respect to the overall reference population)*

| UDA | Competitions with female committee president | Number of competitions with female committee members | | | | | | Female full professors |
|---|---|---|---|---|---|---|---|---|
| | | 0 | 1 | 2 | 3 | 4 | 5 | |
| Mathematics and computer science | 8 (14.0) | 19 (33.3) | 20 (35.1) | 14 (24.6) | 4 (7.0) | - | - | 184 (17.5) |
| Physics | 7 (15.6) | 31 (68.9) | 11 (24.4) | 3 (6.7) | - | - | - | 56 (7.2) |
| Chemistry | 6 (11.1) | 29 (53.7) | 16 (29.6) | 7 (13.0) | 2 (3.7) | - | - | 143 (15.2) |
| Earth sciences | - | 14 (70.0) | 4 (20.0) | - | 2 (10.0) | - | - | 44 (13.0) |
| Biology | 13 (17.6) | 19 (25.7) | 26 (35.1) | 18 (24.3) | 8 (10.8) | 3 (4.1) | - | 416 (28.3) |
| Medicine | 9 (6.0) | 96 (63.6) | 35 (23.2) | 17 (11.3) | 2 (1.3) | 1 (0.7) | - | 287 (11.8) |
| Agricultural and veterinary sciences | 5 (9.8) | 33 (64.7) | 13 (25.5) | 2 (3.9) | 3 (5.9) | - | - | 123 (14.2) |
| Civil engineering and architecture | 12 (14.6) | 40 (48.8) | 29 (35.4) | 8 (9.8) | 3 (3.7) | 1 (1.2) | 1 (1.2) | 153 (13.8) |
| Industrial and information engineering | 10 (7.0) | 106 (74.6) | 28 (19.7) | 8 (5.6) | - | - | - | 97 (5.4) |
| History, philosophy, pedagogy and psychology | 34 (30.6) | 26 (23.4) | 35 (31.5) | 31 (27.9) | 13 (11.7) | 4 (3.6) | 2 (1.8) | 462 (28.8) |
| Economics and statistics | 19 (15.3) | 43 (34.7) | 47 (37.9) | 32 (25.8) | 2 (1.6) | - | - | 309 (17.5) |
| Political and social sciences | 9 (22.5) | 10 (25.0) | 20 (50.0) | 9 (22.5) | - | 1 (2.5) | - | 116 (24.1) |
| Law | 17 (15.3) | 44 (39.6) | 41 (36.9) | 20 (18.0) | 5 (4.5) | 1 (0.9) | - | 327 (17.9) |
| Ancient history, philology, literature and art | 38 (38.0) | 15 (15.0) | 22 (22.0) | 22 (22.0) | 29 (29.0) | 8 (8.0) | 4 (4.0) | 644 (41.8) |
| Total | 187 (16.1) | 525 (45.2) | 347 (29.9) | 191 (16.4) | 73 (6.3) | 19 (1.6) | 7 (0.6) | 3,361 (18.7) |

*Table 2: Female representation in selection committees for academic recruitment (percentages in brackets)*

## 4. The relationships between gender and competition outcomes

In this section we first examine the gender of committee president and members in relation to the expected competition outcomes; then the gender of the president and members in relation to gender of the competition winners.

### 4.1 Relation between the gender of the committee president and members and the competition outcome

The committees' evaluation criteria for candidates were not limited to research performance, thus the definition of expected outcomes is not simple. In our case, on the basis of the available data, research performance is the only indicator we can measure in quantitative terms. As diverse and important as factors other than research performance clearly are, we will in any case assume that a candidate should not win a competition if: i) 50% of the other associate professor candidates have better scientific performance, or ii) if, regardless of the scientific performance of the other candidates, 50% of all Italian assistant professors in the relevant SDS have better performance[9]. From this, in analyzing the relation between the committee president's gender and the competition outcome, we then consider the competition outcome: i) as not expected, if the winners' 2004-2008 FSS is below the median of the distributions for the set of the competition applicants or for all the assistant professors in the same SDS; otherwise ii) as expected. To ensure a robust analysis given these parameters, at this point we further exclude those competitions (46 out of 287) lacking at least a winner and a non-winning participant with at least three years in a faculty position over the 2004-2008 period. Given the exclusions, we reduce the number of competitions observed to 243. Of these there are 106 competitions where the outcomes are as expected, of which 14 had a female committee president, and 137 with non-expected outcomes, of which 7 had a female committee president. The test for association of competitions with expected outcomes and female committee president shows a Pearson chi-square result of 4.96 (Table 3), with a significant p-value of 0.026, and a likelihood-ratio chi-square of 4.94, with significant p-value of 0.026. The odds ratio (OR) results as 2.83, demonstrating a moderate positive association between competitions with expected outcomes and the fact that the committee president is a woman.

In the analysis of the relations between the committee members' gender and competition outcome, we divide the competitions into two groups: those where there are at least three female committee members (president included); and those where there are less than three female committee members. This comparison shows that of the 107 competitions with expected outcomes, six were guided by committees including three or more women. Similarly, of the 137 competitions with non-expected outcomes, six were again guided by committees including three or more women. The Pearson chi-square results as 0.21 (Table 3), with a non-significant p-value at 0.648, and the likelihood-ratio chi-square is 0.21 with a non-significant p-value at 0.649.

We also conduct the same analysis for the competitions guided by committees with two and one female committee members. In both cases the correlations tests result as non significant.

---

[9] Our assumption reflects the regulations laying out the evaluation criteria for the latest national habilitation competitions. Committees were required, among others, to compare the candidate's scores along three bibliometric performance indicators, against the national medians of the distribution for all existing professors in the given SDS.

**4.2 Relation between the gender of the committee president and members and the winners' gender**

In this analysis we examine the association between the gender of the committee president and members and the gender of the competition winners (Table 3, column 3-5). Because research productivity is not involved in this analysis, we can consider the 1,162 total competitions, as well as the subsets of the 654 competition in the hard sciences and the 508 in the remaining disciplines.

In the case of the 1,162 total competitions, both winners are women in 210 competitions, of which 56 had a female committee president. Of the 952 remaining competitions, 131 had a female committee president. The test for association shows a Pearson chi-square result of 21.22, with a significant p-value of 0.000, and a likelihood-ratio chi-square of 19.08, with a significant p-value of 0.000. OR results as 2.28, demonstrating a more than moderate positive association between the fact that the committee president's gender is female and the fact that the competition shows both winners as female.

In the case of the 654 competitions in the hard sciences, both winners are women in 81 competitions, of which 17 had a female committee president. Of the remaining 573 competitions, 54 had a female committee president. The test for association shows a Pearson chi-square result of 9.81, with a significant p-value of 0.002, and a likelihood-ratio chi-square of 8.23, with significant p-value of 0.004. OR results as 9.14, demonstrating a strong positive association between the fact that the committee president's gender is female and the fact that the competition shows both winners as female.

In the case of the 508 competitions in disciplines other than hard sciences, both winners are women in 129 competitions , of which 38 had a female committee president. Of the 379 remaining competitions, 77 had a female committee president. The test for association shows a Pearson chi-square result of 5.37, with a significant p-value of 0.020, and a likelihood-ratio chi-square of 5.15, with significant p-value of 0.023. The OR results as 1.70, demonstrating a modest positive association between the fact that the committee president's gender is female and the fact that the competition shows both winners as female.

We repeat the same tests of association between committee members' gender and the winners' gender for the 1,162 total competitions. The result is a moderate positive association between the case that the committee members' gender is female and the case that the competition shows both winners as female.

In the case of the 654 hard sciences competitions, the test demonstrates a more than moderate positive association between the fact that the committee members' gender is female and the fact that the competition shows both winners as female.

In the case of the 508 competitions in the remaining disciplines the odds ratio results demonstrate more than moderate positive association between the fact that the committee members' gender is female and the fact that the competition shows both winners as female.

|  |  | Winners' gender |  |  |  |
|---|---|---|---|---|---|
|  |  | All competitions (1,162) | Competitions in the hard sciences (654) | Competitions in the other disciplines (508) | Outcome |
| President's gender | Pearson chi-square | 21.22*** | 9.81*** | 5.37** | 4.96** |
|  | Odds ratios | 2.28 | 9.14 | 1.70 | 2.83 |
| Members' gender | Pearson chi-square | 27.23*** | 4.29** | 12.38*** | 0.21 |
|  | Odds ratios | 3.07 | 2.49 | 2.50 | 1.31 |

*Table 3: Analyses of association between committee president gender, winners' gender and outcome*
*Statistical significance: \*p-value <0.10, \*\*p-value <0.05, \*\*\*p-value <0.01*

## 5. Statistical Analysis

We now formulate a statistical model that links the competition outcome to the determinants described below. The dependent variable, the competition outcome, is a Boolean type variable with value of 1 in the case that the applicant wins, or 0 otherwise. The independent variables are eight, including three social proximity variables, two research collaboration variables, one scientific merit variable, and two gender association variables. They are: i) the parental link between applicant and ordinary professors in the same university; ii) the career years that an applicant has spent in the same university and same SDS as the committee president; iii) the career years that an applicant has spent in the same university and the same SDS as other committee members; iv) the percentage of the president's publications coauthored with the candidate; v) the number of other committee members with which the applicant has co-authored publications; vi) as a proxy of scientific merit, the applicant's scientific productivity FSS for the five years 2004-2008 (with citations observed as of 31/12/2011); vii) the agreement between the gender of the applicant and the gender of the committee president; viii) the agreement between the gender of the applicant and the gender of at least three committee members (president included).

As the basis of the statistical model we choose the logistic regression function (rendered linear through the *logit* function), which is particularly suited for modeling dichotomous dependent variables. The statistical model is described:

$$logit(p) = \beta_0 + \beta_1 FSS + \beta_2 NE + \beta_3 CP + \beta_4 CE + \beta_5 PP + \beta_6 PE + \beta_7 SP + \beta_8 SE \quad [2]$$

Where:
$$logit(p) = \log \frac{p(E)}{1-p(E)} \quad [3]$$

$E$ = competition outcome, with a value of 1 if the applicant wins the competition, otherwise 0;
$p(E)$ = probability of event E;
$\beta$ = generic regression coefficient;
$FSS$ = applicant's research productivity over the period 2004-2008, expressed on a 0-100 percentile scale;
$NE$ = value of 1 if the applicant and a full professor in the same university have the same family name; otherwise 0.
$CP$ = applicant's career years in the same university and same SDS as the committee president over the period 2001-2010.

*CE* = applicant's career years in the same university and the same SDS as the other evaluation committee members over the period 2001-2010.

*PP* = percentage of committee president's publications in co-authorship with the candidate over the period 2001-2010.

*PE* = number of other committee members with which the applicant has co-authored publications over the period 2001-2010.

*SP* = value of 1 if the applicant is the same gender as the committee president; otherwise 0.

*SE* = value of 1 if the applicant is the same gender as at least three committee members (president included); otherwise 0.

The evidence from a previous work by Abramo et al. (2014c) suggests that, without distinction for gender, the fundamental determinant of a job candidate's success is not scientific merit, rather the number of years that the candidate has belonged to the same university as the competition committee president. Because the gender of a committee president is a random event, in this work, we carry out the statistical analyses distinguishing the competitions on the basis of the president's gender, to gain an idea of the varying influence of the regressors in the case that the president is a woman rather than a man.

## 5.1 Descriptive statistics

Prior to applying the statistical model we present the descriptive statistics for the variables in Table 7, distinguishing on the basis of committee president's gender. For each variable we show the average, standard deviation (SD) and the maximum value occurring for the competition winners, non-winners and total applicants in the dataset.

| | Female committee president | | | | | | | | |
|---|---|---|---|---|---|---|---|---|---|
| | Winners | | | Non winners | | | Total | | |
| Var. | Avg | SD | Max | Avg | SD | Max | Avg | SD | Max |
| FSS | 75.52 | 21.76 | 100 | 63.67 | 23.01 | 99.70 | 65.84 | 23.19 | 100 |
| NE | 0.00 | 0.00 | 0 | 0.02 | 0.15 | 1 | 0.02 | 0.14 | 1 |
| CP | 4.18 | 4.33 | 10 | 1.55 | 3.11 | 10 | 2.03 | 3.51 | 10 |
| CE | 0.71 | 2.23 | 9 | 1.57 | 4.05 | 21 | 1.41 | 3.79 | 21 |
| PP | 8.19 | 18.86 | 71.88 | 0.19 | 1.14 | 8.33 | 1.66 | 8.62 | 71.88 |
| PE | 0.08 | 0.27 | 1 | 0.03 | 0.17 | 1 | 0.04 | 0.19 | 1 |
| SP | 0.58 | 0.50 | 1 | 0.38 | 0.49 | 1 | 0.41 | 0.49 | 1 |
| SE | 0.55 | 0.50 | 1 | 0.64 | 0.48 | 1 | 0.63 | 0.49 | 1 |
| | Male committee president | | | | | | | | |
| | Winners | | | Non winners | | | Total | | |
| Var. | Avg | SD | Max | Avg | SD | Max | Avg | SD | Max |
| FSS | 67.66 | 23.42 | 100 | 61.49 | 25.27 | 100 | 63.00 | 24.96 | 100 |
| NE | 0.06 | 0.23 | 1 | 0.04 | 0.20 | 1 | 0.04 | 0.20 | 1 |
| CP | 4.27 | 4.38 | 10 | 1.41 | 3.16 | 10 | 2.11 | 3.71 | 10 |
| CE | 1.27 | 3.57 | 20 | 1.25 | 3.27 | 20 | 1.25 | 3.35 | 20 |
| PP | 6.91 | 16.86 | 91.18 | 1.00 | 6.60 | 97.36 | 2.46 | 10.44 | 97.36 |
| PE | 0.13 | 0.38 | 3 | 0.09 | 0.30 | 2 | 0.10 | 0.32 | 3 |
| SP | 0.73 | 0.45 | 1 | 0.70 | 0.46 | 1 | 0.71 | 0.46 | 1 |
| SE | 0.72 | 0.45 | 1 | 0.70 | 0.46 | 1 | 0.71 | 0.46 | 1 |

*Table 4: Descriptive statistics for logistic regression variables*

In the case of competitions where the committee president is a woman, the winners' scientific performance is on average higher than that of non-winners (75.52 for winners versus 63.67 for non-winners). The average number of years that the applicants have worked in the same university as the committee president is 2.03; for winners this figure rises to 4.18 and for non-winners it drops to 1.55. For the set of all applicants, the average number of years spent in the same university as the other committee members is 1.41, compared to 0.71 for the winners and 1.57 for non-winners. Concerning publications, on average the full set of participants contribute to 1.66% of the president's scientific production; winners contribute to 8.19% and non-winners to 0.19%. Concerning gender, 41% of the applicants are of the same gender as the committee president; this percentage rises to 58% in the case of the winners and drops to 38% in the case of the non-winners.

In the case of competitions where the committee president is a man, the winners' scientific performance again averages higher than that of non-winners (67.66 for winners versus 61.49 for non-winners), although the difference is less than in the case where the committee president is a woman. The average number of years that the applicants have worked in the same university as the committee president is 2.11; for winners this figure rises to 4.27 and for non-winners it drops to 1.41. For the set of all applicants and for non-winners, the average number of years spent in the same university as the other evaluators is 1.25, compared to 1.27 for the winners. Concerning publications, on average the full set of participants contribute to 2.46% of the president's scientific production; winners contribute to 6.91% and non-winners to 1.00%. Concerning gender, 71% of the applicants are of the same gender as the committee president; this percentage is 73% in the case of the winners, and 70% in the case of non-winners. Finally, we observe that concerning the variable used for signaling possible cases of nepotism, this systematically assumes a nil value in all the competitions presided by women. However in the competitions presided by men, on average 6% of the winners and 4% of non-winners have the same family name as a full professor in the same university.

**5.2 Correlation analysis**

Table 7 presents the correlations between the regressors, both in the case the committee president is a woman and the committee president is a man.

In the first case, the Pearson correlation analysis indicates that the highest correlation is between SP and SE, at -0.902. The second highest correlation is between PP and E, at 0.359. This is in line with the results from the preceding study by Abramo et al. (2014c). The test of multicollinearity between the variables shows that the average VIF (Variance inflation factor) is 2.29 and that for the variables SP and SE we obtain elevated values of VIF (respectively 5.85 and 5.75), invalidating the hypothesis of independence between the variables. We thus choose to eliminate the variable SE from the model, instead preferring SP, in consideration of the seeming high importance of the committee president's judgment in the final evaluation of the candidates (Abramo et al., 2014c). Excluding the variable SE, the VIF becomes 1.10 with a maximum value of 1.18 (CP). Thus from the Pearson correlation analysis and the test of multicollinearity between the variables it emerges that, excluding SE, the hypothesis of independence between the variables can be considered valid.

|   | \multicolumn{9}{c}{Female committee president} |
|   | E | FSS | NE | CP | CE | PP | PE | SP | SE |
| --- | --- | --- | --- | --- | --- | --- | --- | --- | --- |
| E | 1 | | | | | | | | |
| FSS | 0.198*** | 1 | | | | | | | |
| NE | -0.066 | -0.034 | 1 | | | | | | |
| CP | 0.291*** | 0.008 | -0.081 | 1 | | | | | |
| CE | -0.088 | -0.052 | 0.207*** | -0.186*** | 1 | | | | |
| PP | 0.359*** | -0.012 | -0.027 | 0.339*** | -0.072 | 1 | | | |
| PE | 0.100 | 0.136* | -0.028 | -0.116* | 0.229*** | -0.039 | 1 | | |
| SP | 0.159** | -0.040 | -0.047 | 0.110 | 0.071 | 0.142** | -0.016 | 1 | |
| SE | -0.071 | -0.027 | 0.036 | -0.099 | -0.070 | -0.045 | 0.000 | -0.902*** | 1 |
|   | \multicolumn{9}{c}{Male committee president} |
|   | E | FSS | NE | CP | CE | PP | PE | SP | SE |
| E | 1 | | | | | | | | |
| FSS | 0.107*** | 1 | | | | | | | |
| NE | 0.033 | -0.012 | 1 | | | | | | |
| CP | 0.333*** | -0.004 | -0.008 | 1 | | | | | |
| CE | 0.003 | -0.053** | -0.026 | -0.204*** | 1 | | | | |
| PP | 0.244*** | 0.048** | 0.006 | 0.368*** | -0.079*** | 1 | | | |
| PE | 0.058** | 0.006 | -0.038 | -0.092*** | 0.361*** | -0.037 | 1 | | |
| SP | 0.027 | 0.128*** | -0.002 | 0.036 | -0.041* | 0.007 | -0.027 | 1 | |
| SE | 0.016 | 0.105*** | 0.003 | 0.032 | -0.025 | 0.015 | -0.032 | 0.898*** | 1 |

*Table 5: Correlation among variables*
*Statistical significance: *p-value <0.10, **p-value <0.05, ***p-value <0.01*

In the second case, the Pearson correlation analysis indicates that the highest correlation is between SE and SP, at 0.898. The second highest correlation is between PP and CP, at 0.368. This is in line with what we expect, since scientists in the same university and SDS would tend to cooperate in shared research work. The test of multicollinearity between the variables shows that the average VIF is 1.87 and that for the variables SP and SE we obtain elevated values of VIF (respectively 5.21 and 5.18), invalidating the hypothesis of independence between the variables. Excluding the variable SE, the average VIF is 1.11 and the maximum value is 1.20 (CP and CC). Thus again in the case of male committee presidents, from the Pearson correlation analysis and the test of multicollinearity between the variables, it emerges that the hypothesis of independence between the variables can be considered valid.

**5.3 The logistic regression model**

Continuing from the correlation analysis, in the logistic regression model we again exclude the variable SE, as already mentioned. We also exclude the variable NE, since as we have noted, when competitions are presided by women there are no winners where the variable NE takes a value other than 0. Table 6 presents the logistic regression results predicting the competition outcomes.

In the case of a female committee president, the odds ratio for the competition outcomes (i.e. probability of winning the competition relative to the probability of not winning) is formalized as:

$$\frac{p(E)}{1-p(E)} = \exp(-4.765 + 0.034 * FSS + 0.126 * CP - 0.051 * CE + 0.143 * PP + 1.432 * PE + 0.682 * SP)$$

[4]

The value $e^b$, calculated for each potential explanatory variable, represents OR in Table 6. Where OR equals 1 the associated explanatory variable would have no effect on the dependent variable of "competition outcome". The values calculated for standardized $b$ (last column, Table 6) permit comparison of the effects of the variables measured in different metrics. The data indicate that the factor having the greatest influence on the competition outcomes ($b\text{Std}_{PP}$=1.234, p-value=0.051) seems to be the co-authorship of publications with the committee president (PP). In particular, every percent increase in PP increases the odds ratio by a factor of 1.154. The applicant's scientific productivity (FSS) also has remarkable bearing on the competition results ($b\text{Std}_{FSS}$=0.778; p-value<0.01), with every unit increase in the FSS increasing the odds of success by a factor of 1.034. As well, the number of the applicant's years in the same university and same SDS as the committee president (CP) has notable weight ($b\text{Std}_{CP}$=0.443, p-value=0.027), with every unit increase in the number of career years shared with the president increasing the odds ratio by a factor of 1.135.

| | Female committee president | | | | | |
|---|---|---|---|---|---|---|
| | $b$ | OR | Std Err | Z | p>|z| | $b\text{Std}_X$ |
| FSS | 0.034*** | 1.034 | 0.012 | 2.827 | 0.005 | 0.778 |
| CP | 0.126** | 1.135 | 0.057 | 2.208 | 0.027 | 0.443 |
| CE | -0.051 | 0.950 | 0.073 | -0.704 | 0.482 | -0.194 |
| PP | 0.143* | 1.154 | 0.073 | 1.949 | 0.051 | 1.234 |
| PE | 1.432 | 4.188 | 0.879 | 1.629 | 0.103 | 0.276 |
| SP | 0.682 | 1.977 | 0.421 | 1.619 | 0.106 | 0.337 |
| Constant | -4.765*** | - | 0.987 | -4.830 | 0.000 | |
| | Male committee president | | | | | |
| | $b$ | OR | Std Err | Z | p>|z| | $b\text{Std}_X$ |
| FSS | 0.012*** | 1.012 | 0.003 | 4.591 | 0.000 | 0.295 |
| CP | 0.178*** | 1.198 | 0.016 | 11.343 | 0.000 | 0.669 |
| CE | 0.046** | 1.047 | 0.019 | 2.458 | 0.014 | 0.154 |
| PP | 0.027*** | 1.027 | 0.006 | 4.286 | 0.000 | 0.281 |
| PE | 0.518*** | 1.678 | 0.182 | 2.851 | 0.000 | 0.166 |
| SP | 0.045 | 1.046 | 0.134 | 0.337 | 0.736 | 0.021 |
| Constant | -2.596*** | - | 0.209 | -12.400 | 0.000 | |

*Table 6: Logistic regression results predicting competition outcomes*
*Dependent variable: competition outcome; method of estimation: logistic regression; b = raw coefficient; OR= Odds ratio (exp b); z = z-score for test of b=0; p>|z| = p-value for z-test; bStd$_X$= X standardized coefficient.*
*Statistical significance: \*p-value <0.10, \*\*p-value <0.05, \*\*\*p-value <0.01.*
*Female committee president: Number of observations = 208; LR chi2(6) = 46.40; Prob > chi2 = 0.0000; Log likelihood = -75.6962; Pseudo R2 = 0.2346; mean VIF=1.10*
*Male committee president: Number of observations = 1,771; LR chi2(6) = 245.52; Prob > chi2 = 0.0000; Log likelihood = -864.5317; Pseudo R2 = 0.1243; mean VIF=1.13*

In the case of the male committee president, the odds ratio for the competition outcomes is formalized as:

$$\frac{p(E)}{1-p(E)} = \exp(-2.596 + 0.012 * FSS + 0.178 * CP + 0.046 * CE + 0.027 * PP + 0.518 * PE + 0.045 * SP)$$

[5]

The data indicate that the factor having the greatest influence on the competition outcomes ($b\text{Std}_{CP}$=0.669) seems to be the number of the applicant's years in the same university as the committee president. In particular, every unit increase in the number of career years shared with the president increases the odds ratio by a factor of 1.198. The applicant's scientific productivity (FSS) has less notable weight ($b\text{Std}_{FSS} = 0.295$), with every unit increase in the FSS increasing the odds ratio by a factor of 1.012. Co-authorship of publications with the committee president (PP) has similar bearing on the competition results ($b\text{Std}_{PP}=0.281$), with every percent increase in PP increasing the odds ratio by a factor of 1.027. Shared research work with other committee members has a lesser bearing ($b\text{Std}_{PE} = 0.166$) on competition outcome, however still to a significant level. Similarly, the applicant's career years with the other evaluators ($b\text{Std}_{CE} = 0.154$) also has lesser bearing, and to a lower significance level. In both cases, the fact that the candidate is the same gender as the committee president has no significant bearing on the competition outcome.

## 6. Discussion and conclusions

In a preceding work (Abramo et al., 2014b), we evaluated the efficiency of the selection process for recruitment of university professors in Italy. This analysis revealed several critical issues, particularly concerning unsuccessful candidates who outperformed the competition winners in terms of productivity over the subsequent triennium, as well as a number of competition winners who resulted as totally unproductive. An analysis of the individual competitions showed that almost half of them selected candidates who would go on to achieve below-median productivity in their field of reference over the subsequent period. In a subsequent study (Abramo et al., 2014c), we found that the fundamental determinant of a candidate's success was not his or her scientific merit, rather the number of their years of service in the same university as the committee president. Where the candidate had cooperated in joint research work with the president, the odds ratio also increased significantly. The factors of the years of service and occurrence of joint research with the other committee members had lesser weight. The phenomenon of nepotism, although it occurred, seemed to have a lower impact.

In this paper, we have examined aspects of gender influence on the outcome of academic recruitment. We tried to understand if the gender of the evaluator or applicant could influence the decisional process in the choice among candidates. In addition, the analysis examined the variation of the impact of the determinants that could affect selection procedures when the committee is respectively presided by a man or woman. Among the determinants, other than the scientific merit of the candidate, we considered the effect of favoritism. We identify the potential for favoritism in terms of family links, the history of social proximity, and research collaboration between the candidates and their evaluators, and the propensity to reward candidates of the same gender as the committee president.

Within the limits of all assumptions made, from the analyses of the relationship between gender of the committees and competition expected outcome, we observe a moderate positive association between competitions with expected outcomes and the fact the committee president is a woman. This positive association is weaker and non-significant if we consider the prevailing gender within the committee (gender of at least three out of the five committee members), rather than that of the president. Further, in the case of competitions in the hard sciences, we encounter a strong positive association between the fact that the committee president is a woman and the fact that the competition shows both winners as female. This association becomes weaker in the case of selection competitions in the other disciplines. Similarly, the association between the fact of having at least three committee members of female gender and the fact of having female winners is more than modest.

From the statistical analysis, aimed at assessing the different weights of factors determining competition outcomes on the basis of committee's president gender, we observe that in the case that the committee president is a woman, the factor having the greatest influence on the competition outcomes seems to be the co-authorship of publications with the committee president, followed by the applicant's scientific productivity and the number of the applicant's years in the same university as the committee president. In the case that the committee president is a man, the factor having the greatest influence on the competition outcomes seems to be the years together with the candidate in the same university, followed by research productivity and shared publications. In the case of committees presided by women, scientific merit has a greater influence on the success of a candidate that in those presided by men. The fact that the candidate is the same gender as the committee's president seems to have no significant bearing on the competition outcome.

These results suggest that compared to their male colleagues, the female professors called to evaluate competitions give more importance to merit. If the productivity metric is a fair proxy of the candidate merit, then female committee presidents seem to assure more expected outcomes, resulting more "fair" in their assessment. This suggests a policy remedy in placing women in greater number of presiding positions to increase the quality of selected pool and the representation of women in these positions. Thus the recommendations for the Italian policy maker are to favor greater female representation on the competition committees, in particular in the committee president role. The question is whether this recommendation can also be extended to other countries, given the differing conclusions from other studies. While holding that our measure of scientific performance is much more precise and robust than those used in preceding studies, we believe that the different results could only in part be attributed to this factor. In our consideration there are two country-specific factors that are much more important: i) the intensity of competition in the higher education sector; and ii) the levels of corruption and favoritism in society. In nations where the higher education system is highly competitive, it is to be expected that phenomena of discrimination and favoritism would be much less, under parity of other conditions, compared to nations with scarce competition among universities. In fact in nations with more competitive systems, non-efficient recruitment would in all probability have negative consequences, both direct and indirect, for the committee members themselves. In these countries, gender differences in the behavior of selection committee should be more rare and for the most part associated with socio-cultural reasons. In nations where the higher education system is scarcely competitive but the levels of corruption in society are low, one might witness few cases of gender differences in judgment. In nations where the low intensity of competition among universities is further associated with high levels of corruption and favoritism, resorting to female committee presidents could instead be revealed as a means to improve fairness of judgment in recruitment selection. Further investigations could indeed concern cross-country comparisons. In conclusion, the characteristics of the context could explain the different results arriving from the various studies on the theme, meaning that the policy recommendations also cannot be independent of the context where they are to be applied. Alternative approaches to study gender influences on the outcome of academic recruitment could be considered as well. The selection process in fact entails social interaction among the members of the committee. So far results in the literature are mainly based on indirect reconstructions of review processes (van Arensbergen et al 2014). Direct observation of the group dynamics occurring during the review and decision making processes is therefore desirable.